# Correlation between crystal structure and superconductivity in LaO$_{0.5}$F$_{0.5}$BiS$_2$

J. Kajitani[1], K. Deguchi[2], A. Omachi[1], T. Hiroi[1], Y. Takano[2], H. Takatsu[3], H. Kadowaki[3], O. Miura[1], Y. Mizuguchi[1,2]

1. Department of Electrical and Electronic Engineering, Tokyo Metropolitan University, 1-1, Minami-Osawa, Hachioji, Tokyo 192-0397, Japan
2. National Institute for Material Science, 1-2-1, Sengen, Tsukuba, Ibaraki 305-8565, Japan
3. Department of Physics, Tokyo Metropolitan University, 1-1, Minami-Osawa, Hachioji, Tokyo 192-0397, Japan

Abstract

Correlation between crystal structure and superconducting properties of the BiS$_2$-based superconductor LaO$_{0.5}$F$_{0.5}$BiS$_2$ was investigated. We have prepared LaO$_{0.5}$F$_{0.5}$BiS$_2$ polycrystalline samples with various lattice constants. It was found that the annealing the sample under high pressure generated uniaxial strain along the $c$ axis. Further, the highly-strained sample showed higher superconducting properties. We concluded that the uniaxial strain along the $c$ axis was positively linked with the enhancement of superconductivity in the LaO$_{1-x}$F$_x$BiS$_2$ system.

*Introduction*

The layered crystal structure is one of the notable stages to explore new superconductors with a high transition temperature ($T_c$) and to discuss the mechanism of unconventional superconductivity. Recently, several kinds of $BiS_2$-based superconductors such as $Bi_4O_4S_3$ and $REO_{1-x}F_xBiS_2$ (RE = rare earth), which possess the $BiS_2$ conduction layers, have been discovered [1-6]. The crystal structure of the $BiS_2$-based superconductors is composed of an alternate stacking of conduction layers and blocking layers. This is quite similar to the crystal structure of the cuprates and the Fe-based superconductors [7,8]. Therefore, it is expected that new $BiS_2$-based superconductors with high $T_c$ would be discovered. The parent phase of the $BiS_2$-based family is basically an insulator with $Bi^{3+}$, and superconductivity in $BiS_2$-based materials is emerged by electron doping to the Bi-6p orbital within the $BiS_2$ layers [1,9]. Among them, $LaO_{0.5}F_{0.5}BiS_2$ has the highest $T_c$. Although the $T_c$ (onset) of the As-grown (solid-state-reacted) $LaO_{0.5}F_{0.5}BiS_2$ is only 3 K, it can be enhanced to 10.6 K by applying annealing under high pressure. It was found that the crystal structure obviously changed after high-pressure annealing, indicating that the superconducting properties depended on not only the carrier concentration but also the changes in crystal structure [2, 10, 11]. However, the correlation between the superconducting properties and the crystal structure remains to be clarified. Here, we show that the superconducting properties of $LaO_{0.5}F_{0.5}BiS_2$ can be tuned with changing the lattice constants. In particular, the uniaxial strain is important for the appearance of higher $T_c$ in $LaO_{0.5}F_{0.5}BiS_2$.

*Experimental*

Polycrystalline samples of $LaO_{0.5}F_{0.5}BiS_2$ were prepared by solid-state reaction using powders of $La_2S_3$ (99.9%), $Bi_2O_3$ (99.9%), $BiF_3$ (99.9%), $Bi_2S_3$ and grains of Bi (99.99%). The $Bi_2S_3$ powder was prepared by reacting Bi (99.99%) and S (99.99%) grains at 500 ºC in an evacuated quartz tube. The starting materials with a nominal composition of $LaO_{0.5}F_{0.5}BiS_2$ were well-mixed, pressed into pellets, sealed into an evacuated quartz tube, and heated at 700 ºC for 10h; here we call this sample "As-grown" sample.

As reported in refs. 2 and 11, the $T_c$ of $LaO_{0.5}F_{0.5}BiS_2$ is enhanced by annealing the as-grown sample under high pressure. Therefore, the obtained sample was annealed at 600 ºC under high pressure (HP) of 2 GPa using a cubic-anvil high-pressure synthesis instrument. Here we call this product "HP" sample.

To investigate intermediate specimens, we annealed the HP sample in an evacuated quartz tube at various temperatures because annealing could generally reduce the strain. Hence, to realize tuning of crystal structure, the obtained "HP" sample was cut into several pieces, separately sealed into an evacuated quartz tube and annealed at 300, 500 and 700 °C for 10 hours, where the heating rate was fixed at 200 °C / h, respectively. The samples were furnace-cooled down to room temperature. These samples are called with the annealing temperature, for example, "300 °C sample" for the sample annealed at 300 °C.

The powder X-ray diffraction experiments were carried out using a Rigaku SmartLab powder diffractometer equipped with a CuK$_{\alpha 1}$ monochrometer. To compare XRD patterns of various samples precisely, the standard Si powder was mixed with the samples, and used as the reference of the calibration of the peak shift. The obtained X-ray diffraction patterns were analyzed using the Rietveld method. The temperature dependence of electrical resistivity from 300 to 2 K was measured using the four-terminal method. The DC magnetic susceptibility was measured using a superconducting quantum interference device (SQUID) magnetometer after zero-field cooling (ZFC) with an applied field of ~5 Oe. The transition temperature ($T_c^{mag}$) was defined to be a temperature at which the signal of diamagnetic susceptibility reached 1 % of -1/4π. The microstructures were investigated using a scanning electron microscope (SEM).

## Results and discussion

Figure 1(a) and 1(b) show powder X-ray diffraction patterns of the As-grown sample and the 700 °C sample, respectively. These diffraction patterns were reasonably fitted by parameters of the space group of *P*4/*nmm*, which were analyzed by using RIETAN-FP [12]. The obtained lattice constants were $a$ = 4.0710(1) Å, $c$ = 13.3495(3) Å for the As-grown sample, $a$ = 4.0718(1) Å, $c$ = 13.3780(3) Å for the 700 °C sample. The $R$ factors of the Rietveld refinement were obtained as $R_{wp}$ = 13.3 %, $R_e$ = 7.1 % ($S$ = 1.8) for the As-grown sample, $R_{wp}$ = 10.2 %, $R_e$ = 5.7 % ($S$ = 1.8) for the 700 °C sample, respectively. The resulting structure parameters are listed in Tabs. 1 and 2. There is no large change in crystal structure between the As-grown sample and the 700 °C sample, indicating both the annealing under HP (2 GPa, 600 °C) and the post annealing at 700 °C in evacuated tube do not change the space group and do not produce impurities. For the HP, 300, 500 °C samples, we did not obtain reliable refinements, because of the peak broadening. However, clear shifts of peak positions in the X-ray diffraction profiles were observed. This indicates the lattice is strained by applying HP annealing, but returned by the post annealing the sample at 700 ºC in an evacuated tube.

To investigate the characteristics of the generated strain after the HP annealing, the enlargement of X-ray profiles around the (004) peaks and the (200) peaks for all the samples are shown in Figs. 2(a) and 2(b), respectively. We find that the peaks of the HP sample are obviously broader than that of the As-grown sample. Then, the peaks become sharper with applying post annealing at 300, 500 and 700 ºC. Finally, the X-ray profile of the 700 ºC sample becomes comparable to that of the As-grown sample as confirmed in Fig 1. The (004) peak shifts to higher angles, which indicates large contraction of the $c$ axis with applying HP annealing. In contrast, the (200) peak slightly shifts to lower angles after the HP annealing, indicating slight expansion of the $a$ axis. Due to the asymmetric peaks, the reliable estimation of the lattice constants for the HP, 300 and 500 ºC samples are difficult. However, to discuss the change in lattice constants qualitatively, the lattice constants were calculated using the peak positions at the maximum intensity in the (004) and (200) peaks, and summarized in the insets of Figs. 2(a) and 2(b). In the calculation of the lattice constants, the peak-shift parameters estimated by the Rietveld refinement for the As-grown sample were used. It is clear that the length of the $a$ axis decreases and the length of the $c$ axis increases with applying HP annealing. In addition, we find that the peak symmetry is different in between the ($00l$) and ($h00$) peaks. the broadening of the (004) peak is relatively large, and the peaks are asymmetric. In contrast, the broadening of the (200) peak is small, and the broadened peaks are almost symmetric. These facts suggest that the $c$ axis is easily compressed, and the uniaxial strain (compression) along the $c$ axis is generated after HP annealing in the $LaO_{0.5}F_{0.5}BiS_2$ system.

Figure 3 (a) shows the temperature dependence of electrical resistivity for all the samples. For all the samples, resistivity increases with decreasing temperature; namely, transport properties are semiconducting-like even though the electron carriers are introduced in the conduction layers upon F substitution at the O site. The resistivity at 300 K for the As-grown sample is the highest, and the resistivity at 300 K becomes lower by applying HP annealing. The application of the post annealing at 500 and 700 ºC enhances the value of resistivity at 300 K. To compare the $T_c$ for all the samples, the temperature dependence of resistivity was normalized at 15 K and displayed in Fig. 3 (b). The HP sample shows the highest transition temperature of $T_c^{zero}$ = 7.0 K. With applying post annealing at 300, 500 and 700 ºC, $T_c$ decreases with increasing post-annealing temperature. Finally, the $T_c$ for the 700 ºC sample almost corresponds with that of the As-grown sample, which is consistent with the X-ray diffraction results. In this regard, the $T_c$ in the $LaO_{0.5}F_{0.5}BiS_2$ system is governed by the change in crystal structure such as lattice strain.

Figure 4 shows the temperature dependence of magnetic susceptibility (ZFC) for all the samples. The As-grown sample shows weak superconducting (diamagnetic) signal with $T_c^{mag}$ = 2.4 K. The HP sample shows the highest $T_c^{mag}$ of 6.9 K with a large shielding volume fraction. With increasing post-annealing temperature, the $T_c$ decreases as observed in the resistivity measurements, and the shielding volume fraction decreases in the 500 and 700 ºC sample. Thus, the uniaxial strain generated by applying HP annealing affects not only $T_c$ but also the shielding property.

The change in microstructure due to the generation of strain was investigated by the SEM observation. Figure 5 (a) and (b) display the SEM images for the As-grown and the HP samples, respectively. For the As-grown sample, grains with a size of several μm are observed. In Fig. 5 (b), it is found that the grain size for the HP sample is smaller than that for the As-grown sample. Further, the surface of the HP sample seems to be amorphous-like. These changes should be resulted from the application of the HP annealing, in other words, the generation of uniaxial strain.

Finally, we summarize and discuss the evolution of superconductivity and structural changes in the $LaO_{0.5}F_{0.5}BiS_2$ system. On the basis of the X-ray diffraction, resistivity and susceptibility measurements for the $LaO_{0.5}F_{0.5}BiS_2$ samples prepared by various annealing conditions, we conclude that the uniaxial compression along to the $c$ axis is essential for both the appearance of bulk superconductivity and higher $T_c$. In fact, this scenario is consistent with the theoretical study by Suzuki *et al.*, which suggests that the band structure depends on the distance between La in the blocking layer and S in the $BiS_2$ layers [13]. The uniaxial compression along to the $c$ axis should affect the band structure. Hence, the superconducting properties might be strongly enhanced, probably due to the enhancement of density of states near the Fermi level. The other possibility to explain the evolution of superconductivity is improvement of the localization of electron carriers. As mentioned above, this system shows anomalous metallic behavior. Furthermore, theoretical studies predict that a charge-density-wave stability exists in the $BiS_2$-based family [14,15]. Therefore, it is possible that the uniaxial strain suppress an ordered state such as a charge-density-wave state. To obtain more detailed information on a relationship between crystal structure and superconducting properties in the $BiS_2$-based superconductor, crystal structural analysis using single crystals under high pressure is needed.

### Conclusion

The correlation between crystal structure and superconducting properties of the $BiS_2$-based superconductor $LaO_{0.5}F_{0.5}BiS_2$ has been investigated. We have synthesized $LaO_{0.5}F_{0.5}BiS_2$ polycrystalline samples with various annealing conditions up to 3 steps. The HP annealing generates uniaxial strain along the $c$ axis. The generated strain is returned to the initial state of the As-grown sample by annealing the sample in an evacuated quartz tube at 700 ºC. The highest superconducting properties, $T_c$ and shielding fraction, are observed in the HP sample, and the superconducting properties is degraded by reducing the uniaxial strain. On the basis of those results, we conclude that the enhancement of the superconducting properties in $LaO_{1-x}F_xBiS_2$ by applying post-annealing under high pressure is caused by the generation of the uniaxial strain along the $c$ axis.

Acknowledgement

This work was partly supported by a Grant-in-Aid for Young Scientists (A) and The Thermal and Electric Energy Technology Foundation.

Table 1. Crystallographic data of the As-grown samples.

| Site | $x$ | $y$ | $z$ | Occ. | $U_{iso}$ (Å$^2$) |
|---|---|---|---|---|---|
| La | 0.5 | 0 | 0.1008(1) | 1.0 | 3.8×10$^{-3}$ (fixed) |
| Bi | 0.5 | 0 | 0.6215(1) | 1.0 | 3.8×10$^{-3}$ (fixed) |
| S1 | 0.5 | 0 | 0.3752(4) | 1.0 | 6.3×10$^{-3}$ (fixed) |
| S2 | 0.5 | 0 | 0.8114(4) | 1.0 | 6.3×10$^{-3}$ (fixed) |
| O/F | 0 | 0 | 0 | 0.5/0.5 | 3.8×10$^{-3}$ (fixed) |

Table 2. Crystallographic data of the 700 ºC samples.

| Site | $x$ | $y$ | $z$ | Occ. | $U_{iso}$ (Å$^2$) |
|---|---|---|---|---|---|
| La | 0.5 | 0 | 0.1005(1) | 1.0 | 3.8×10$^{-3}$ (fixed) |
| Bi | 0.5 | 0 | 0.6218(1) | 1.0 | 3.8×10$^{-3}$ (fixed) |
| S1 | 0.5 | 0 | 0.3750(3) | 1.0 | 6.3×10$^{-3}$ (fixed) |
| S2 | 0.5 | 0 | 0.8111(4) | 1.0 | 6.3×10$^{-3}$ (fixed) |
| O/F | 0 | 0 | 0 | 0.5/0.5 | 3.8×10$^{-3}$ (fixed) |

Figure captions

Fig. 1. X-ray diffraction patterns of (a) the as-grown sample and (b) the 700 °C sample. Observed and refined data are shown by crosses and solid curves, respectively. The difference between the data and the model is plotted by the dashed curves in the lower part. Vertical bars represent positions of the Bragg reflections.

Fig. 2. (a) Enlargement of X-ray profile near the (004) peaks for all the samples. The inset shows the annealing condition dependence of lattice constant $c$ estimated using the peak position of the (004) peak. (b) Enlargement of X-ray profile near the (200) peaks for all the samples. The inset shows the annealing condition dependence of lattice constant $a$ estimated using the peak position of the (200) peak. The lattice constants estimated using the Rietveld refinement are also plotted in the insets.

Fig. 3. (a) Temperature dependence of resistivity for all the samples. (b) Temperature dependence of resistivity for all the samples.

Fig. 4. Temperature dependence of magnetic susceptibility for all the samples.

Fig. 5. (a) SEM images for the As-grown sample. (b) SEM images for the HP sample.

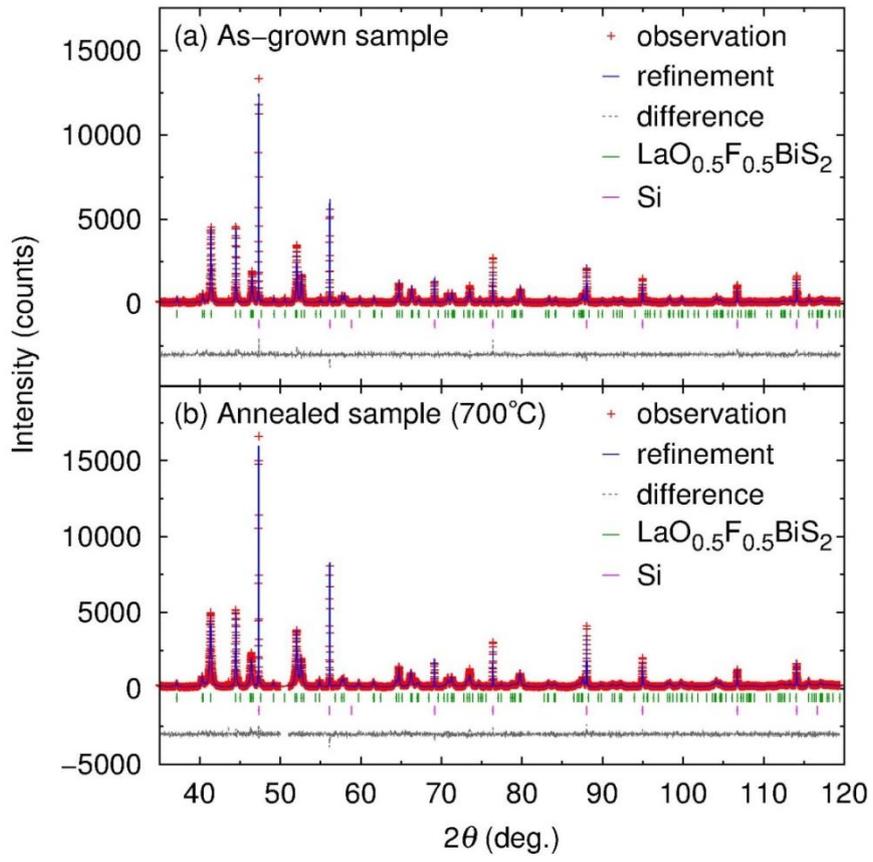

Fig. 2

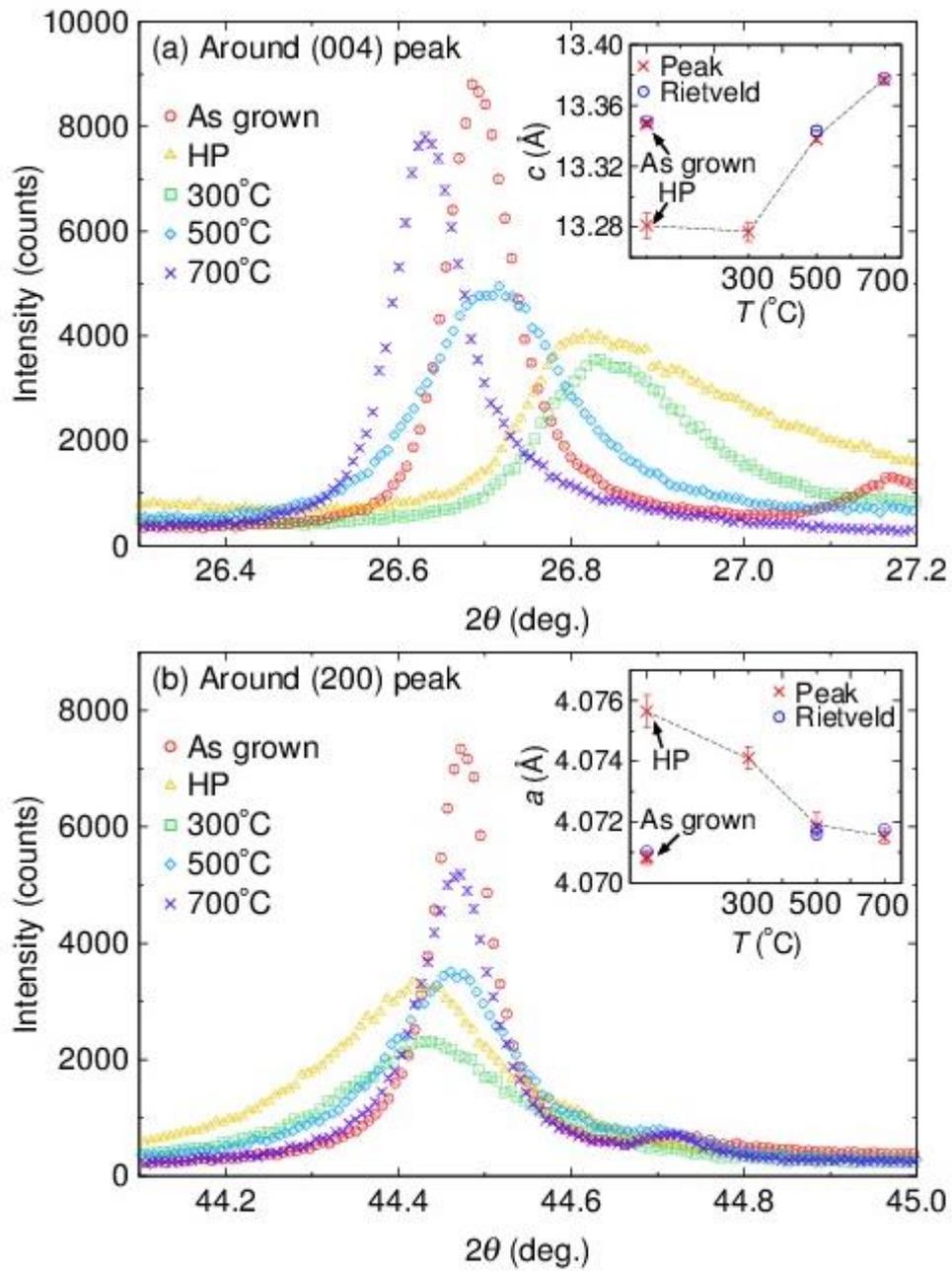

Fig. 3

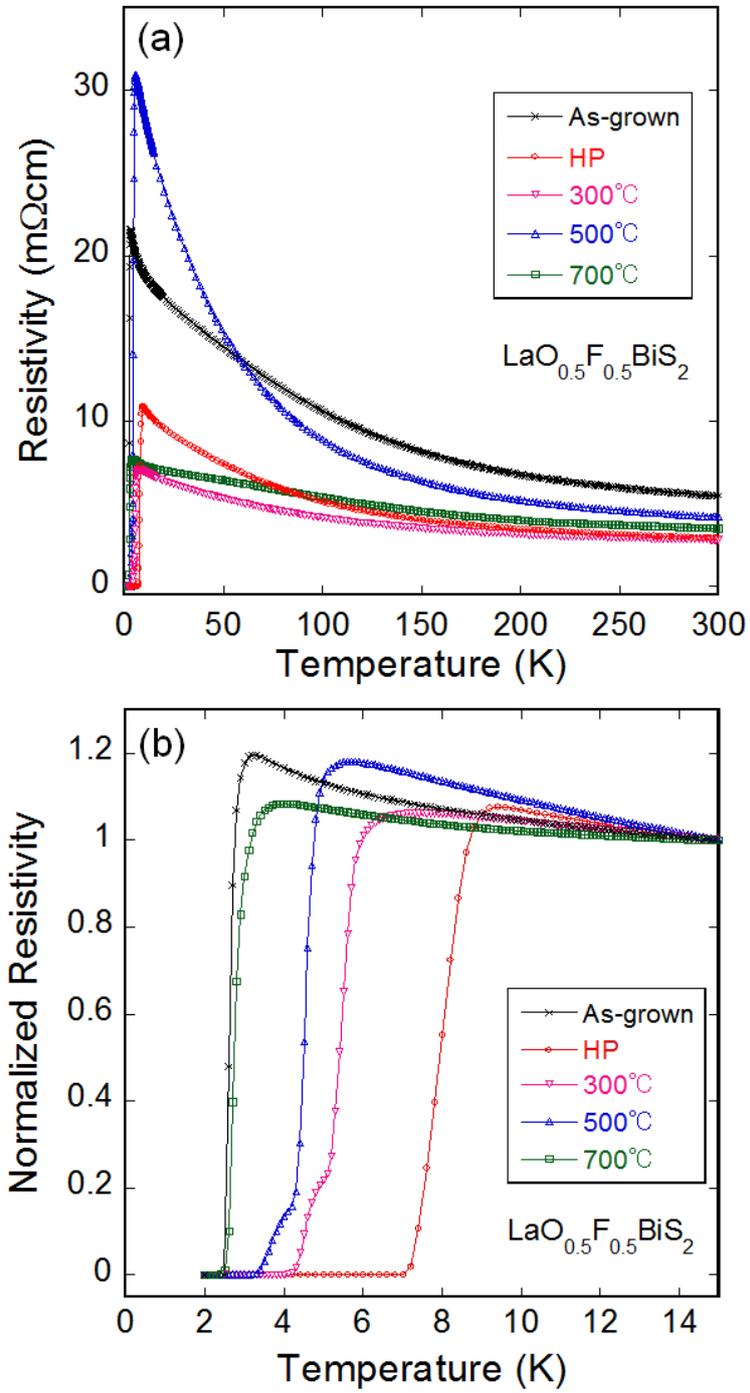

Fig. 4

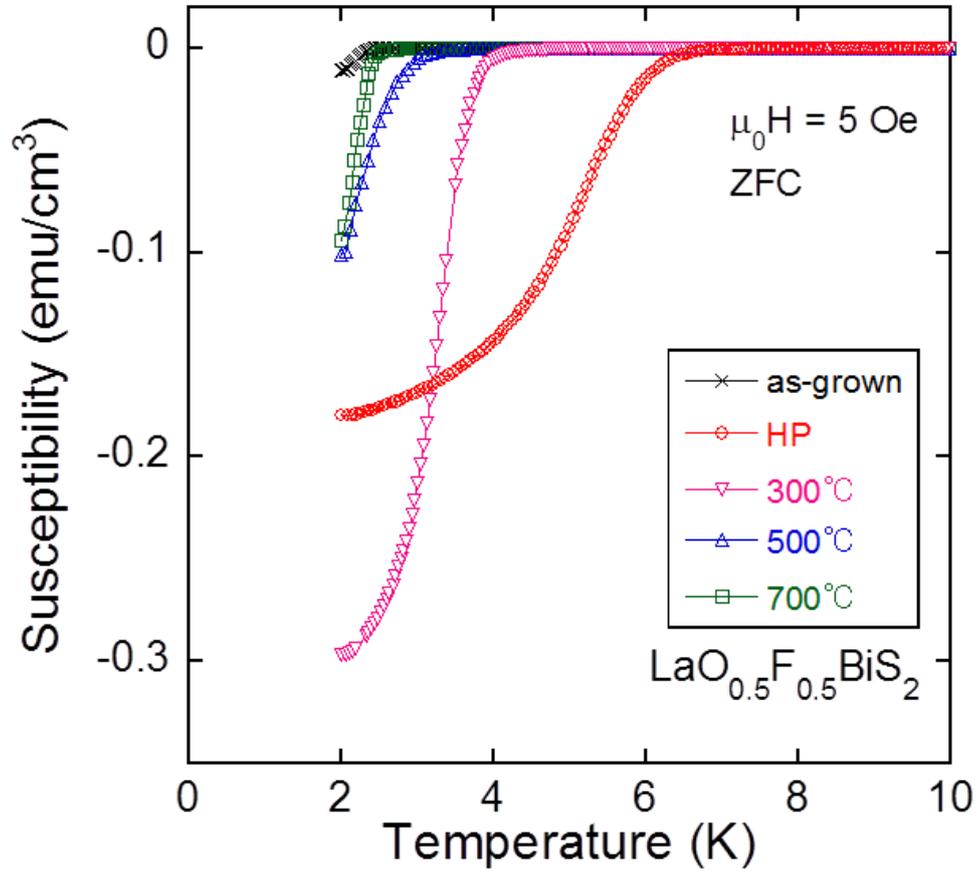

Fig. 5

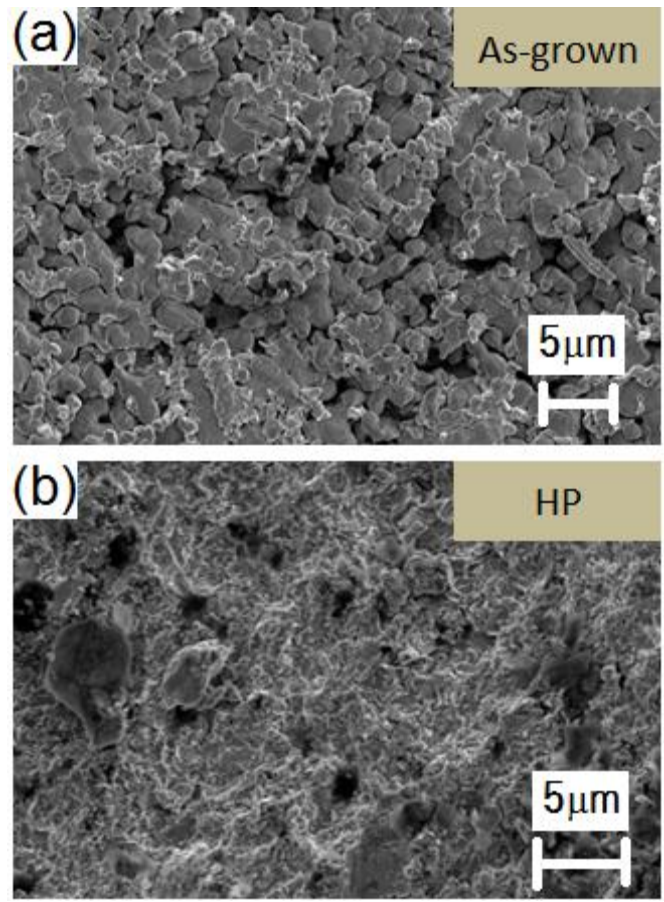